\documentclass[twocolumn,showpacs,preprintnumbers,
amsmath,amssymb,groupedaddress]{revtex4}

\usepackage{graphicx}

\begin{document}

\title
{ The ground state of clean and defected graphene: 
Coulomb interactions of massless Dirac fermions, 
pair-distribution functions and  spin-polarized phases.
}

\author
{ M.W.C. Dharma-wardana
}
\email[Email address:\ ]{chandre.dharma-wardana@nrc.ca}
\affiliation{
National Research Council of Canada, Ottawa, Canada. K1A 0R6\\
}

\date{29 june 2006}
%
%
\begin{abstract}
First-principles density functional calculations for graphene and defected
graphene are used to examine when
the quasi-2D electrons near the Fermi energy in graphene could be
represented by massless fermions obeying a Dirac-Weyl (DW) equation.
The DW model is found to be inapplicable to defected graphene 
containing even $\sim$3\% vacancies or N substitution.
However, the DW
model holds in the presence of weakly adsorbed molecular layers.
The possibility of spin-polarized phases (SPP) of DW-massless fermions
in pure graphene is considered. The exchange energy is evaluated
from the analytic pair-distribution functions as well as in $k$-space.
The kinetic energy enhancement of the sipn-polarized phase nearly
 cancels the exchange enhancement, and
the correlation energy plays a dominant residual role. The correlation
energies are estimated via a model four-component 2D electron fluid whose
Coulomb coupling matches that of graphene.
While SPPs appear with exchange only,
the inclusion of correlations suppresses them in ideal graphene. 
\end{abstract}
%
\maketitle
\section{Introduction.}
\begin {figure}
\includegraphics*[width=8.0cm, height=9.0cm]{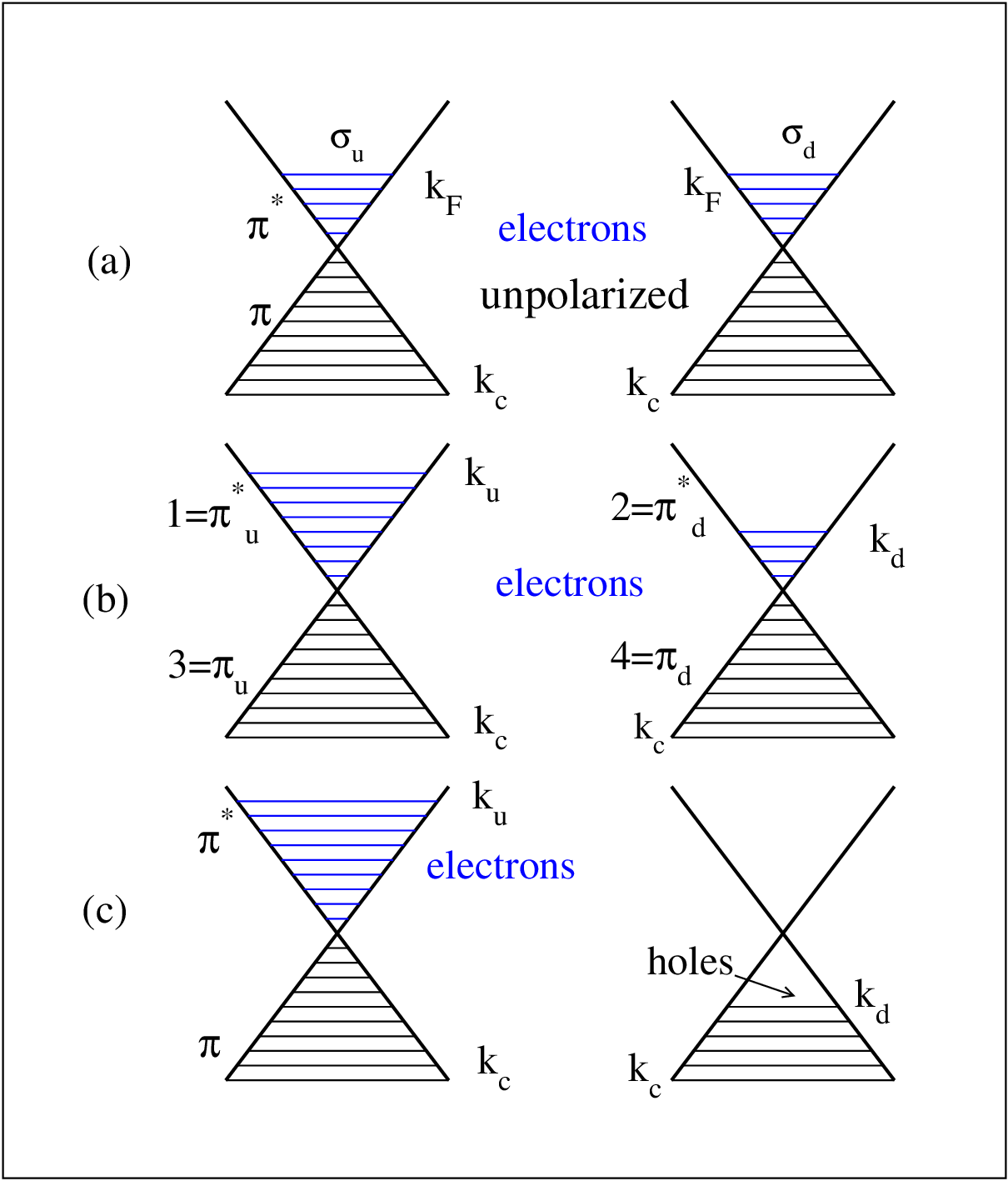}
\caption{
(Color online)Linear dispersion bands near a {\bf K} point where the
$\pi^*$ and $\pi$ bands cross. In (a) we show a doped unpolarized
system with equal occupation of the up-spin ($\sigma_u$) and
down-spin states. In (b) the polarized system has only electron
carriers. In (c) both electron and hole carries occur. This is
the only possibility for spin polarization 
if doping is zero ($k_F=0$ and $k_u=k_d$).
}
\label{figcones}
\end{figure}

Graphene and related materials (e.g, nanotubes, fullerenes)
 have become a mine of  novel technologies and new 
 horizons in physics\cite{geim}. These include
 cosmological models on
 honeycomb branes,
 superconductivity on bi-partite lattices\cite{nagamatsu},
 nanotubes\cite{cdw,lorin}, Hubbard models\cite{tosatti},
 spin-phase
transitions\cite{pgcn}, nanostructures\cite{takis}
 and other
aspects of strongly correlated electrons\cite{khvesh}.
The carbon atoms in graphene form a quasi-two-dimensional (Q2D) honeycomb
 lattice and contribute one electron per carbon to form an unusual 2D electron
system (2DES) with a massless Fermions obeying a Dirac-Weyl (DW)
equation near
 the Fermi points\cite{wallace,dvm,ando}.
The hexagonal Brillouin zone has two inequivalent
points {\bf K}=$(1/3,1/\surd{3})$ and {\bf K$^\prime$}=$(-1/3,1/\surd{3})$,
in units of $2\pi/a_0$, where $a_0$ is the lattice constant. 
The simplest tight-binding model with nearest-neighbour hopping $t$
is sufficient to describe the
 valence
and conduction bands ($\pi$ and $\pi^*$) 
 near the {\bf K, K$^\prime$} points, i.e, at the Fermi energy $E_F$,
where the bandgap is zero.
The Dirac-Weyl 2D electron system (DW-2DES)  
is nominally ``half-filled'', with the $\pi^*$ band unoccupied,
(see Fig.~\ref{figcones}) and has spin and
valley degeneracies, with a Berry phase associated with the
 valley index\cite{ando}.

The above picture assumes a perfect 2-D sheet of C atoms
in a honeycomb lattice held in place by the $\sigma$-bonding
structure of the C framework. In practice, since defects are favoured
by the entropy term in the free energy, some  carbon atoms may be missing,
forming vacancies; they may also be
substituted with other defect atoms. The surface itself may be covered
with adsorbed gases. Hence the nature of the density of states (DOS) near
the Fermi energy in systems with vacancies and substituted atoms needs
to be considered\cite{nieminen,lee}. Removing a carbon atom effectively
removes four valance electrons from the system, and the resulting vacancy may
or may not lead to a magnetic, conducting or insulating ground state.
 Replacing a C atom by, say, a
nitrogen atom provides five valance electrons. We find that at typical
 concentrations of 3\% or more, the Dirac-Weyl picture fails, and the
Fermi energy moves into a bandgap or to regions with a high density
 of states. However,
if less extreme situations are considered (e.g, adsorbed gases on graphene),
the DW picture is found to hold true.

%

%
The Fermi liquid found in metals and
semi-conductor interfaces is characterized by the Wigner-Seitz
radius $r_s$ of the sphere (or disk in 2D) which contains one electron.
When expressed in atomic units (involving the effective mass $m^*$
of the electrons), $r_s$ becomes the ratio of the potential
energy to the kinetic energy, i.e, a measure of {\it the strength of the
Coulomb interaction}. Hence $r_s<1$ provides a regime where
Fermi-liquid perturbation theory is strictly valid.
The vanishing of the density of states and the effective mass of the
electrons in DW-2DES near the Fermi points implies that the 
strict Fermi-liquid picture breaks down in perfect graphene.

The DW-2DES can be doped to contain carrier electrons (in the $\pi^*$ band),
or holes as well (in the $\pi$ band), and provides a rich system
which retains the massless-fermion character, as long as the material 
is not strongly modified. 
 However, electron-electron 
interactions in the DW-2DES may modify
 the $\pi$ and $\pi^*$ bands,
lift the sublattice (valley) degeneracies, or stabilize
spin-polarized phases (SPP) in preference to the unpolarized state, if the
doping level or the strength of the Coulomb interaction could be varied.
The effect of electron-electron interactions has been most extensively studied
in Fermi-liquid-like electron systems found in GaAs/AlAs interfaces
 or Si/SiO$_2$-inversion layers.
The SPP in such GaAs/AlAs-based  2DES, predicted to appear at low coupling ($r_s\sim 2-4$)
if perturbation methods are used, get pushed to
high coupling ($r_s\sim ~26$) if non-perturbative theory (NPT)
were used\cite{attac,prl3}. The two-valley
2DES does {\it not} show a SPP in NPT, unlike for one-valley systems,
presumably because of the preponderance of the (three
times as many) direct Coulomb interactions
over the  exchange interactions\cite{2valley}.
The exchange and correlation energy $E_{xc}$ in the 4-component Si/SiO$_2$
electron system were calculated
in Ref.\onlinecite{2valley}
using the classical-map hyper-netted-chain (CHNC)
technique, accurately recovering the Quantum Monte-Carlo (QMC) results
in every coupling regimes\cite{conti}. CHNC provides the
pair-distribution functions (PDFs) $g_{ij}$ as a function of the coupling
strength.
Then $E_{xc}$ is evaluated
via a coupling constant integration, providing a fully non-local,
transparent approach.
The method has been successfully applied to the 2DES\cite{prl2},
 3DES\cite{prl1}, the 2-valley 2DES in SiO$_2$ interfaces\cite{2valley},
the thick quasi-2DES in HIGFETS\cite{q2des}, and the
electron-proton system\cite{hyd}. However, the full non-local treatment of
exchange and correlation in graphene involves an $8\times8$ matrix of
two-component PDFs because of the spin, and valley indices as well as the
presence of $\pi,\pi^*$ bands. Hence in this study we first consider the
exchange energies via an analytic evaluation of the non-interacting
PDFs. The correlation energies are estimated by appealing to our 
results for the spin-polarized four-component two-valley 2-D
electron system (2v-2DES) of Si-MOSFETS. 
 The non-interacting PDFs of the DW-2DES,
$\mathcal G^0_{ij}(r)$
 involve two components, the first being
a Bessel function as in the ordinary 2DES, while a second,
 associated with the cosine of the angle of 
e-e scattering, involves products of Bessel and Struve functions.
The exchange energy enhancement of the sipn-polarized phase nearly
 cancels the kinetic energy enhancement,
 implying that
the correlation energy plays the dominant residual role.
We find that while there are stable SPPs in an exchange-only approach,
including the correlation energy using the 2v-2DES data
stabilizes the 
graphene-2DES in the unpolarized state. This conclusion is not surprising since
the Coulomb coupling strength in graphene is $\sim$ 2-3, and
no SSPS are found in electron liquids at such low coupling, except as
artifacts of low-order theories.

In the following section we present first-principles calculations
for pure graphene, graphene with $\sim$ 3\%, and $\sim$ 12 \% vacancy
concentrations, and show that the DW-model is inapplicable to such systems.
We also consider systems with N-substitution as the lattice distortion effects
are smaller here. Nevertheless, even here the DW-model does not seem
to be applicable. However, if we consider a graphene layer having a 
metastable sheath of N$_2$ molecules adsorbed  on it, with no disruption of
the $\sigma$-bonding network,  the DW-model does remain applicable.
Thus, having established the limits of the DW-model, we proceed to
examine the exchange and correlation effects among DW-fermions, and
show that a stable spin phase transition is found only in ``exchange only''
models which neglect electron correlation effects. 
\section{Density-functional calculations of graphene systems}
Simple tight-binding models(TBM) can be trusted only if they are validated by
more detailed calculations  and experiments. While TBM can be successfully
exploited within a limited energy window for pure graphene, the effect of
vacancies and lattice substituants etc., needs detailed consideration. A
vacancy removes 4 valence electrons, distorts the bond lengths
and angles around the vacancy, creating
pentagonal networks and compensating larger networks, localizing electrons
near the vacancy and changing the structure of the electronic density
of states (DOS). The bond-lengths and the network structure is better 
preserved with N substitution. Here an extra electron is added to the
graphene system for every N substitution.
Attempting to treat such effects using tight-binding methods augmented by, say,
$T$-matrix theory etc., to account for impurity effects are well known to be
strongly model dependent. Nevertheless, some insight has already
been gained  from short-ranged scatterer models where lattice relaxation
and other very important issues are not handled\cite{pgc-v,shon,palee}
However, density-functional
theory (DFT) has an excellent track record in just such problems where
electronic and ionic energy minimization can be carried out until the
Hellman-Feynman forces on atoms around the vacancy or the substituant are
reduced to zero. There  are  a number of such DFT calculations already 
in the literature\cite{nieminen,lee}, mainly concerned with
 energetics and bonding.
Here we examine the bands and DOS of these systems, with   
an eye on the limits of validity of the DW-2DES model.
      
We have used the Vienna {\it ab initio} simulation package (VASP)\cite{vasp}
which implements a spin-density functional periodic plane-wave basis
calculation. The projected augmented wave (PAW) 
pseudopotentials\cite{vasp} have been used for C and N. The C pseudopotential
has already been used in several graphene-type calculations 
(e.g., Ref.~\onlinecite{nieminen}).
The N pseudopotential was
also further tested by a study of the N$_2$ dimer where an equilibrium
bond distance of 1.11 Angstroms was obtained, in good agreement with
other DFT calculations as well as results from 
detailed configuration interaction studies
etc.\cite{ahlrichs}, where a value of 1.095 \AA$\,$
 is reached.
The $\sim$ 3\% and $\sim$ 12\% vacancy calculations were done with
32-atom and 8-atom graphene-like unit cells. These systems are thus {\it not} truly 
disordered, but provide reference densities of states which acquire
smearing when some disorder is introduced by sampling other configurations
using larger simulation cells\cite{iyak-cdw}.
 When vacancies are introduced into
graphene, the large stresses are relieved by
the neighbours (C-atoms near the vacancy) moving
towards the vacancy. The distortion persists to at least the
third neighbours, and generates a bond-length distribution varying from
about 1.398\AA$\;$ to 1.45 \AA.$\,$ If the vacancies are replaced by N substitution,
the structural distortions are smaller but the changes in the DOS can be
equally drastic, as we show below. 

\begin{figure}
\includegraphics*[width=8.0cm,height=10.5cm]{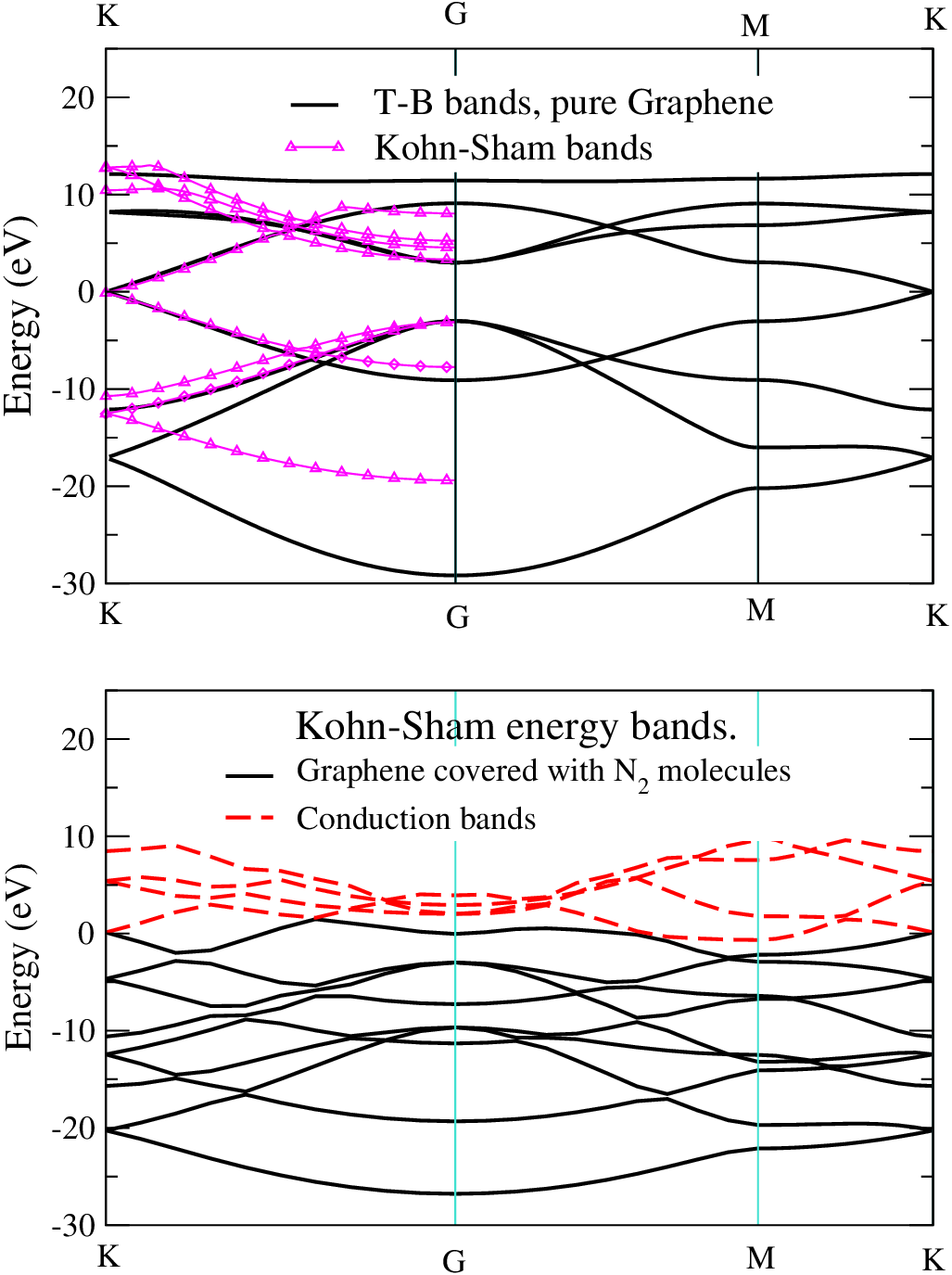}
\caption
{ (Color online) Band structure of pure graphene (top panel),
and that of fully N$_2$ covered graphene. The Fermi energy is
at E$+F$=0. Conduction bands are shown in red. The bands obtained
from a standard $sp^3$ tight-binding scheme are compared with the DFT bands
for graphene in the top panel.
  }
\label{c2n2b}
\end{figure}

Yuchen Ma et al.\cite{nieminen} have found that N$_2$ molecules may form
a metastable layer near carbon nanotubes and graphene. 
Iyakutti et al\cite{iyak-cdw} have also found similar stabilization, where
the N$_2$ layer is less stable than if it were at infinity, but held in
place by an energy barrier. 
Although this is a 'weakly adsorbed' state, the interaction energy between the
graphene layer and the N$_2$ layer is about as strong as between two
graphene sheets. Hence we have looked at the band-structure and DOS of such
a fully N$_2$ covered graphene sheet as well. Here the Dirac-Weyl behaviour is
preserved. In Fig.~\ref{c2n2b} we show the band structure of pure 
graphene (top panel), and also a
graphene sheet with a layer of N$_2$ molecules, with each N$_2$
aligned on every K\'{e}kul\'{e}-bond
position (bottom panel). The sheet of N$_2$ molecules is positioned
about 3\AA$\,$ above the graphene sheet. The interaction energy per carbon (or per N) is
about 0.2 eV. The regime
of D-W linear dispersion around the {\bf K} point is reduced from that of pure graphene.
This is in contrast to the interaction between two graphene sheets, where the
bands near the {\bf K}-point become parabolic\cite{peeters}.

\begin{figure}
\includegraphics*[width=8.0cm,height=10.5cm]{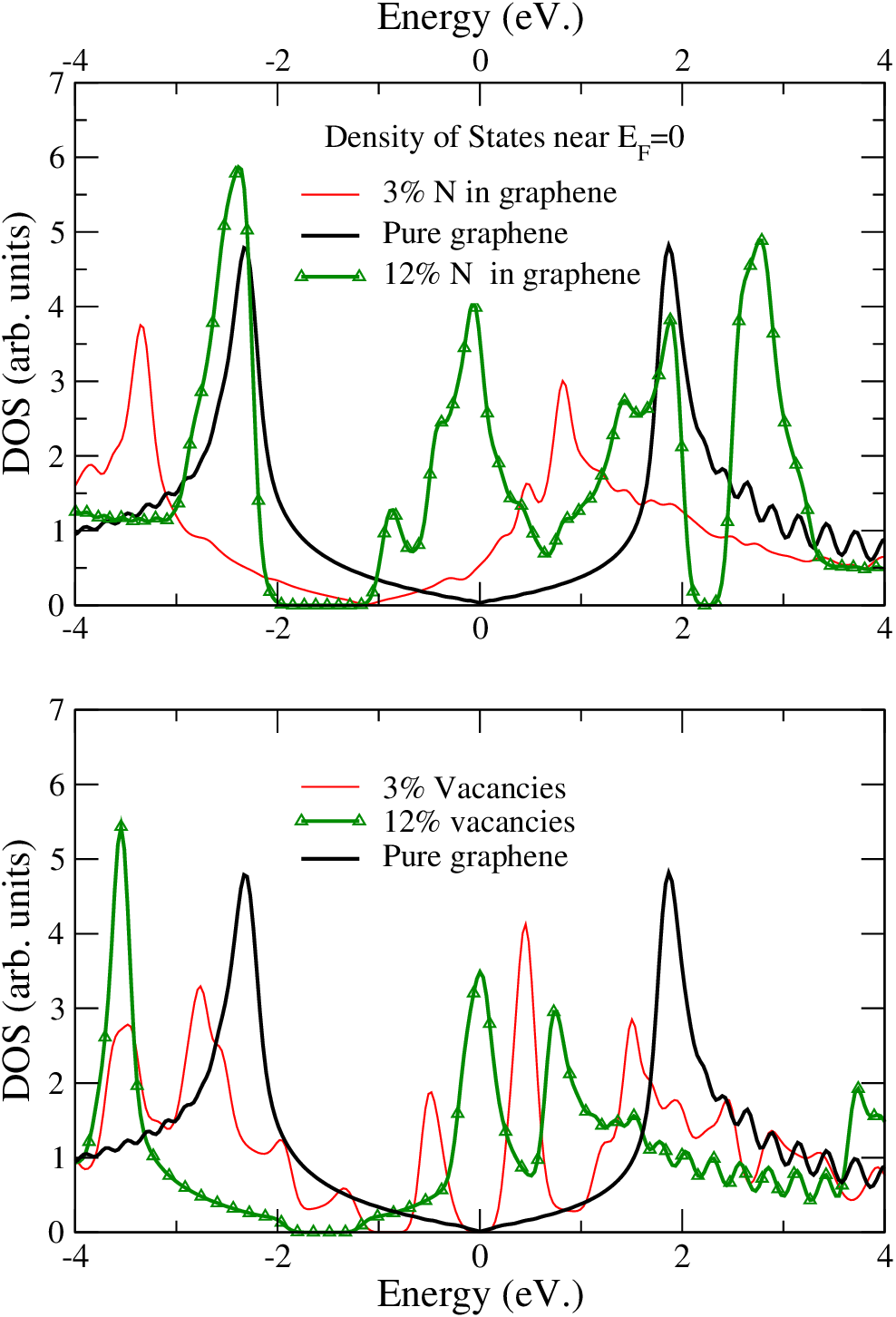}
\caption
{(Color online)DOS of pure graphene compared with $\sim3\%$ and $\sim$12\% concentrations
of substituted N atoms(top panel), calculated using a
periodic substitution model.
The Fermi energy is set to zero. Bottom panel shows the effect of
$\sim 3\%,12$\% vacancies in graphene.
}
\label{grvn}
\end{figure}

However, when vacancies or N atoms are introduced into the
graphene network at relatively significant concentrations (e.g., 3\%-12 \%), the
DOS modifies and shifts to produce a large density of states near the Fermi
energy $E_F$. This is shown in the top panel of Fig.~\ref{grvn}. The calculation
is for an ordered system with 1/8 or 1/32 alloying of vacancies or N, and may be used as
reference points for a consideration of a disordered system. The calculations
used exchange-correlation functionals based on the Ceperley-Alder type\cite{ceperley},
 and no
spin polarized states are expected for a conductive system with an enhance DOS, as seen
in the figure. On the other hand, the existence of such an enhanced
DOS near $E_F$ in graphene with 12\% vacancies, would be of importance to
possible superconductivity of these systems.

The electronic density of states for a 
lower density vacancy system, i.e., with one vacancy per 32 carbon atoms,
is shown in the lower panel. Here the system {\it acquires a gap} near $E_F$,
and the material is an insulator.
If the vacancy concentration is further decreased, we
expect the energy gap to slowly decrease and recover the Dirac-Weyl model
in the limit of pure graphene. But, as seen from the DOS at 12\% vacancies, we
see that as the vacancy concentration is increased from 3\%, the energy gap
closes and $E_F$ positions to a high-DOS region. Meanwhile, an energy gap opens 
about 1 eV below the Fermi energy. Thus we see that a {\it vacancy induced
metal-insulator transition} is possible (see however, Ref.~\onlinecite{altland}).
 This picture becomes considerably
less sharp if the vacancies are not considered to form a periodic array. In any
case, our conclusion that the Dirac-Weyl model is inapplicable even at 3\% vacancy
concentrations probably remains valid. The observations of the quantum Hall effect
and other signatures of the DW model clearly indicate that good low-vacancy
regions of graphene foils are the subject of these experiments. Another aspect
of low-density vacancies or substituents in graphene is the issue of spin-polarized
ground states. The reported results depend on the size and edge
structure of finite sheet fragments\cite{lee}, or the possibility of further
C-adatom adsorption on N-substituted sites\cite{nieminen}.
 The calculations are sensitive
to the treatment of exchange and correlation, as in
 first-principles theories of magnetism in transition-metal oxides.
 Hence a reliable discussion of
magnetism in graphene-like systems would require an assessment of
Coulomb interactions in Dirac-Weyl electrons.
Hence  we revert to the main object
of the present study, viz., the nature of exchange and correlation effects in
pure graphene.    
%

%
\section{The 2DES in Graphene near the Fermi points.}
The band structure of graphene near the {\bf K} points, i.e., close
to the Fermi energy, are displayed in Fig.~\ref{c2n2b}.
The kinetic energy near
the {\bf K} points is given by a Dirac-Weyl Hamiltonian of the form: 
\begin{equation}
\label{hamil}
H_k=V_F(p_x\tau_z\sigma_x+p_y\sigma_y)
\end{equation}
Here $\tau_z=\pm 1$
defines the degenerate valleys, and $\sigma_x,\, \sigma_y$ denote
 the $x$ and $y$ Pauli
matrices that act in the space of the two atoms in each unit cell.
The $\pi,\pi^*$ bands of
spin and valley degenerate states (Fig.~\ref{figcones})
show a linear dispersion $E=\pm V_F\hbar k$.
This form requires
a cutoff momentum $K_c$ such that the number of states in the
Brillouin zone is conserved. That is, if $A_0$ is the 
area per carbon, then 
\begin{equation}
\label{kcutoff}
K_c^2=4\pi (1/A_0)
\end{equation}.
 The electron density
$N_c$ at half-filling is $1/A_0$, with $A_0=a_0^2\surd{3}/2$,
since one $\pi$ electron of arbitrary spin is
provided by each carbon atom. 
The Fermi velocity $V_F=ta_0\surd{3}/2$ is
thus the slope of the linear dispersion, with  $V_F\sim$ 5.5 eV\AA.
If the DW-2DES is embedded in a medium with dielectric
constant $\epsilon_0$, then we define
\begin{equation}
\label{g0def}
g^0=\frac{e^2/\epsilon_0}{\hbar V_F}=\frac{e^2}{\epsilon_0a_0}
/(t\surd{3}/2).
\end{equation}
This is the ratio of a typical Coulomb energy 
to the hopping energy and hence is usually taken as the Coulomb coupling
constant of the DW-2DES.  
This ratio plays the same role as the $r_s$ parameter
in electron-gas theory of nonrelativistic finite-mass fermions, and
is a measure of the strength of the Coulomb interaction.
The usual $r_s$ is not available for
DW-2DES since the effective mass $m^*$ is zero and there is no effective
Bohr radius. The coupling constant $g^0$ is {\it maximized} if $\epsilon_0$
is unity, and consistent with this case
we assume $g^0=2.7$, $V_F=5.39$ eV\AA$\,$, with $e^2/\epsilon_0$=14.4,
for our DW-2DES studies. 

The 4-component-eigenfunction envelopes of the kinetic energy term are
made up of two-component functions $U=(b,e^{i\phi_k})$,
$U'=(e^{i\phi_k,b})$ and  $O=(0,0)$
where $\phi_k$ is the angle of the vector $\vec{k}$ in the 2-D plane.
Thus
\begin{eqnarray}
F^{\bf K}_{b,\vec{k}}(r)&=&(2A)^{1/2}(U,O)_T\chi_\sigma\\
F^{\bf K^\prime}_{b,\vec{k}}(r)&=&(2A)^{1/2}(O,U')_T\chi_\sigma
\end{eqnarray}
Here $b=\pm 1$ is a $\pi^*,\pi$ band index, $(\cdots)_T$ indicates the
transpose, and $\chi_\sigma$ is the spin function.
Then, using $v=1,2$ as a valley index, the Coulomb interaction
term in the Hamiltonian
may be written in the form:
\begin{eqnarray}
\label{hamilt}
\nonumber
H_I&=&\frac{1}{8A}\sum_{v_i,b_i,\sigma_i}\sum_{{\bf k,p,q}}V_q
\left[b_1b_4e^{i\{\phi^*({\bf k})-\phi({\bf k}+\bf{q})\}}+1\right]\\
\nonumber
 & &\times\left[b_2b_3e^{i\{\phi^*({\bf p})-\phi({\bf
 p}+\bf{q})\}}+1\right]\times\\
 & &a^+_{{\bf k},v_1,b_1,\sigma_1}a^+_{{\bf p}+{\bf q},v_2,b_2,\sigma_1}
a_{{\bf p},v_2,b_3,\sigma_2}a_{{\bf k}+{\bf q},v_1,b_4,\sigma_2}
\end{eqnarray}
Here $a^+,a$ are electron creation and annihilation operators and
$V_q=2\pi e^2/(\epsilon_0 q)$ is the 2D Coulomb interaction. The
phase factors introduce a novel $\cos(\theta)$ contribution where
$\theta$ is the scattering angle, not found in the usual jellium-2DES.
The resulting form of the exchange energy per Carbon is:
\begin{eqnarray}
\label{exchange-en}
E_{x}/E_u&=&-\frac{A_0g^0/k_c}{(2\pi)^2}\frac{1}{4}\sum_{b_1,b_2,\sigma}
\int_0^{2\pi} d\theta dk dp \\
\nonumber
 & &\times kp\frac{1+b_1b_2\cos(\theta)}{|{\bf k}-{\bf p}|}
n_{b_1,\sigma}(k)n_{b_2,\sigma}(p)
\end{eqnarray}
In the above we have included the intrinsic coupling
 constant $g^0$ and the
energy unit $E_u=V_Fk_c$ in the expressions.
 Here $k_c=K_c/\surd{2}=\surd(4\pi n_c)$ is
based on the electron density per spin species, $n_c=N_c/2=1/(2A_0)$.
The above form of the exchange energy
can be reduced to an evaluation of a few elliptic integrals\cite{pgcn}.
The normal ``half-filled'' DW-2DES can be doped with electrons or
holes; but it is easy to show that symmetry enables us to limit to
one type of doping. However, given a system, 
with an areal density of $N_\delta$
dopant electrons per valley, with $n_\delta=N_\delta/2$
per spin, the carriers in the spin-polarized
system could be electrons only,
 or both electrons and holes,
as shown in Fig.~\ref{figcones} for the $\pi^*$ and $\pi$ bands at
one {\bf K} point. The intrinsic system with $n_\delta=0$ can  be
an unpolarized state, or spin-polarized state with
electrons {\it and} holes. Such {\it purely exchange driven} systems have
been studied by Peres et al.\cite{pgcn}, while the correlations
effects have not been considered.
Here we evaluate the exchange energy $E_x$
from from the non-interacting
PDFs, and include the correlation energy $E_c$ estimated from the 2v-2DES
with the same coupling strength ($r_s=g^0$) and spin-polarization.
\begin {figure}
\includegraphics*[width=7.0cm, height=8.0cm]{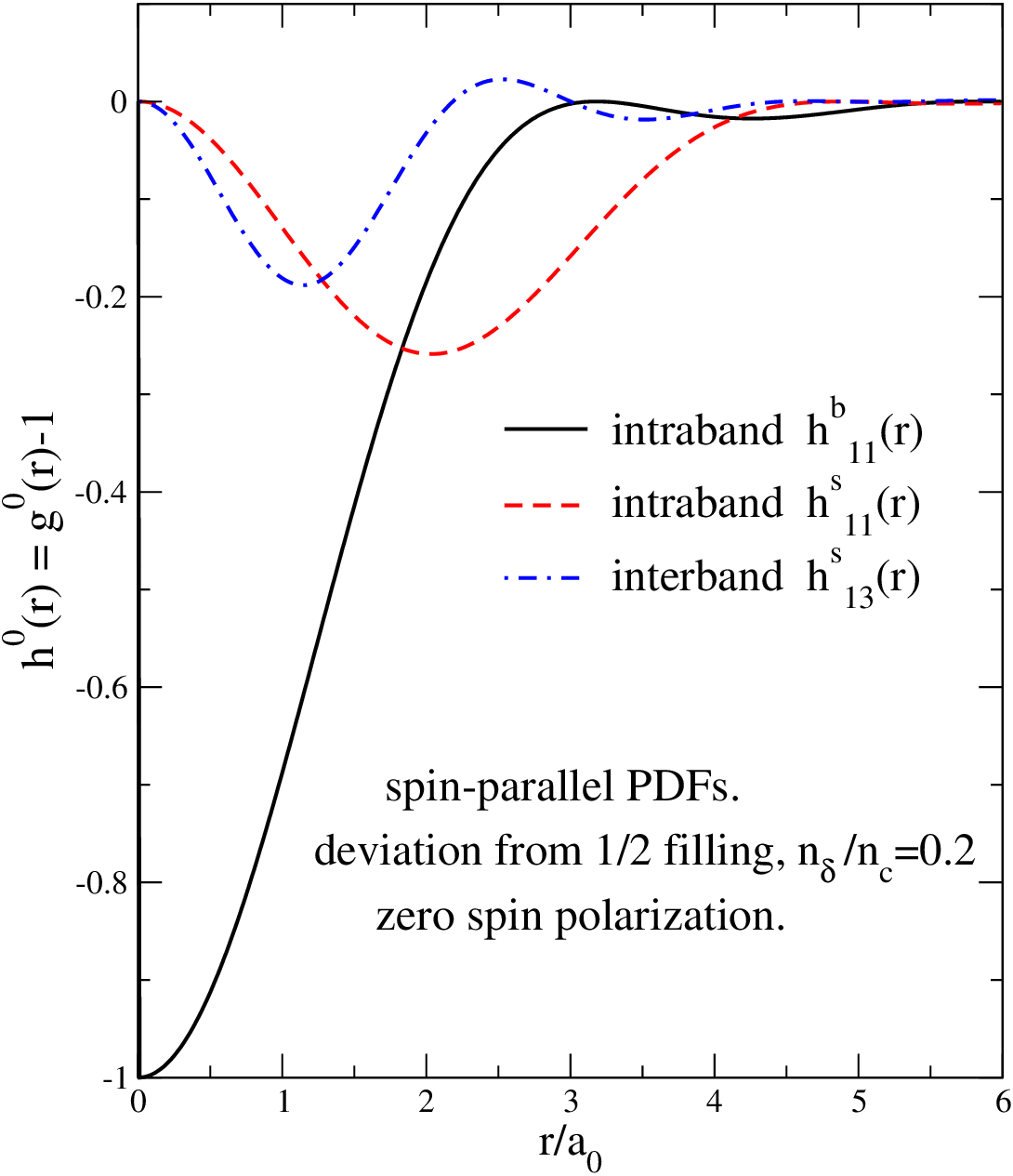}
\caption
{(Color online)The Bessel-like and Struve-like non-interacting,
 parallel-spin PCFs
$h^b(r)$ and $h^s(r)$ for the unpolarized doped system. The
bands are numbers as in Fig.~\ref{figcones}(b). The  anti-parallel
non-interacting PCFs are zero. The lattice constant $a_0=2.47$\AA.
}
\label{figgr}
\end{figure}

\subsection{Electron Pair-distribution functions of the Dirac-Weyl 2DES}
Although we are dealing with an intrinsically four-component
system (2-valleys, 2-spin states), as seen from Fig.~\ref{figcones},
we need to consider the redistribution
of electrons and holes among the $\pi^*$ and $\pi$ bands when
comparing the energy of spin-polarized states with the
corresponding unpolarized
state. The exchange energy is a consequence of the
anti-symmetry of the Slater determinant made up of
non-interacting eigenfunctions. Thus only the
non-interacting PDFs are needed to calculate
the exchange energy. They are also the spring-board
for calculating the interacting PDFS via the CHNC
method.

From Fig.~\ref{figcones} we see that the
e-e interactions at a given valley can be constructed from
(i) interactions with 
a $\pi^*\sigma_u$ band of up-spin electrons of density $n_u$,
 filled to $k_u$,
 (ii) a $\pi^*\sigma_d$
 set of electrons 
or a $\pi\sigma_d$ spin-down holes, of density $n_d$,
filled up to $k_d$ 
(iii)the $\pi\sigma_u$ band, with electron density $n_c$,
 filled to $k_c$, and 
(iv)the $\pi\sigma_d$ band, density $n_c$, filled to $k_c$
There will also be similar intervalley terms.  
Each term in this 4$\times$4 matrix, denoted by $\mathcal G_{ij}(r)$
where $i$, $j=1\cdots4$,
 will have two components
associated with those in $U'$ and $U=(b,e^{i\phi_k})$.
Thus $\mathcal G_{ij}(r)=g^b_{ij}(r),g^s_{ij}(r)$, where the
superfixes `$b,s$' indicate that the noninteracting forms are
Bessel-function like, and Struve-function like, respectively. These
components will be denoted by a superfix $c=b,s$. The
Struve form arises from the $\cos(\theta)$ terms in
the Coulomb interaction.
 The
numbering scheme of the matrix is shown in Fig.~\ref{figcones}(b).
We denote the pair-correlation functions (PCFs)
$\mathcal H_{ij}(r)=\mathcal G_{ij}(r)-1$, or the components
by $h^c_{ij}(r)=g^c_{ij}(r)-1$. The non-interacting forms are
indicated by a superscript zero. Thus we have:
\begin{eqnarray}
\nonumber
h^{0,b}_{ij}(r)&=&-(n_in_j)^{-1}\int_0^{k_i}\frac{d{\bf k_1}}{(2\pi)^2}
\int_0^{k_j}\frac{d{\bf k_2}}{(2\pi)^2} 
e^{i({\bf k}_1-{\bf k}_2)\cdot{\bf r}}\\ 
 &=&-\frac{2}{k_ir}J_1(k_ir)\frac{2}{k_jr}J_1(k_jr) \\
 \nonumber
h^{0,s}_{ij}(r)&=&-(n_in_j)^{-1}\int_0^{k_i}\frac{d{\bf k_1}}{(2\pi)^2}
\int_0^{k_j}\frac{d{\bf k_2}}{(2\pi)^2}\\
\nonumber
 & &\cos(\theta_1-\theta_2)
 e^{i({\bf k}_1-{\bf k}_2)\cdot{\bf r}}\\
\nonumber 
&=&-\frac{\pi}{k_ir}\frac{\pi}{k_jr}[J_0H_1-J_1H_0]_i
[J_0H_1-H_0J_1]_j 
\end{eqnarray}
Here $J_0,J_1$ are Bessel functions, while $H_0$ and $H_1$
are Struve $H$-functions. Also, in $[J_0H_1-J_1H_0]_i$ the
functions are evaluated at the argument $k_ir$. That is,
\begin{equation}
[J_0H_1-J_1H_0]_i=J_0(rk_i)H_1(rk_i)-J_1(rk_i)H_0(rk_i)
\end{equation}
 The wavevectors
$k_i=\surd{(4\pi n_i)}$ for each component $i$, of density $n_i$.
We show typical noninteracting PCFs for a doped, unpolarized case 
as in, Fig.~\ref{figcones}(a), with the doping fraction $n_\delta/n_c=0.2$.
In CHNC, the exchange-hole is mapped {\it exactly} into
a classical Coulomb fluid using the Lado procedure\cite{prl1}. 
The figure shows that the exchange-hole is strongly reduced by the
presence of the $\cos(\theta)$ term which has been averaged into the
Struve-like PCFs $h^s(r)$. When the Coulomb interaction is included,
the $\cos(\theta)$ term has a similar mitigating effect and
exchange-correlation in the DW-2DES is considerably weaker than in the
corresponding 2-valley 2DES. The CHNC calculation for the
2v-2DES for the conditions stipulated in Fig.~\ref{figgr} show that
the correlation energy is only about a third of the exchange energy.
This motivates our use of the 2v-2DES for the correlation
energy, while the $E_x$ is exactly evaluated. 
\subsection{The kinetic and exchange energies.}
\begin {figure}
\includegraphics*[width=7.0cm, height=8.0cm]{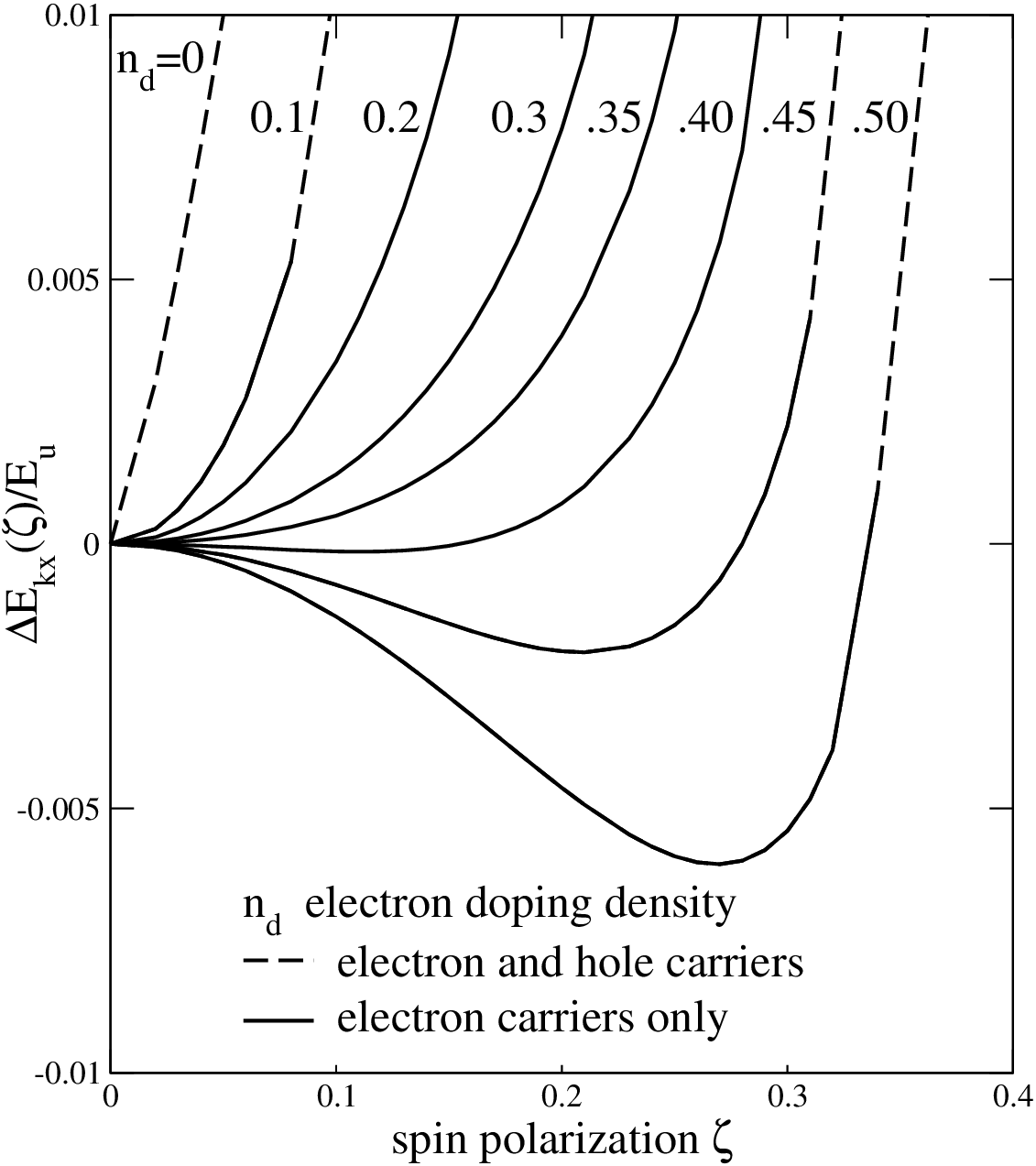}
\caption
{(Color online)Upper panel-
the energy difference $\Delta E_{kx}$, i.e.,
K.E+exchange,
between the polarized and
unpolarized phases, in units of $E_u=V_Fk_c$, as a function
of the spin polarization $\zeta$ and the dopent density $n_d$.
Electron-carrier systems, Fig.~\ref{figcones}(b) are more stable
than electron-hole systems, and show stable spin-polarized states.
However, addition of the correlation energy (see Fig.~\ref{jelcorr})
makes the unpolarized state the most stable phase. 
}
\label{figex}
\end{figure}

When the doping per valley is $N_\delta=2n_\delta$, the total number
 of electrons per valley is
$N_t=N_c+N_\delta$. Also, using the $i=1,2,3,4$ notation of
Fig.~\ref{figcones}(b), we set $n_1=n_u$, $n_2=n_d$, $n_3=n_4=n_c$. Hence the
spin polarization $s=n_u-b_dn_d$, where the band index $b_d=-1$
for holes. The degree of spin-polarization $\zeta=s/N_t$.
The composition fractions, inclusive of the valley index v=1,2 are
 $x_{vi}=n_i/2N_t$. We note that $k_F=\surd{(2\pi n_\delta)}$,
 $k_u=\surd{\{2\pi(n_\delta+s)\}}$, $k_d=\surd{\{2\pi|n_\delta-s|\}}$.
The exchange energy $E_x(n_\delta,\zeta)$ can be written
as:
\begin{eqnarray}
\label{exceq}
 &&E_{x}(n_\delta,\zeta)/N_t= (N_t/2)\int \frac {2\pi r
dr}{r}\\
\nonumber
&&\times\sum_{ij}x_{vi}x_{vj}[\mathcal G^0_{v,v,ij}(r)-1]
\end{eqnarray}
 It is
understood that the Struve-like component in $\mathcal G_{v,v,ij}$
where $v$ labels the valleys, is summed
with the appropriate $b_ib_j$ band $\pm$ factors.
Only a sum over the components in
one valley is needed in evaluating the exchange.
The above formula can be made more explicit, by
introducing the $b_ib_j$ band $\pm$ factors in each case.
Thus the total kinetic and exchange energy $E_{k\,x}$= K.E+$E_x$,
for the case (b) of Fig.~\ref{figcones}
can be written in terms of $n_F, n_u, n_d$
and $n_c$ as in Eq.~\ref{exceq}, or in terms of
 $k_F,k_u,k_d,k_c$ and $A_0$ as:
\begin{eqnarray}
\label{exeq}
&&E_{kx}(\zeta)=\frac{A_0}{6\pi}\hbar V_F(k_u^3+k_d^3)-\\
 \nonumber
&&\frac{A_0}{(2\pi)^2}(g^0\hbar V_F/4)(\pi/2)
[k_u^4 \mathcal H_{11}(r)+ 
k_d^4 \mathcal H_{22}(r)+\\
\nonumber
&&2k_c^2\{k_u^2\mathcal H_{13}(r)+
k_d^2\mathcal H_{14}(r)\}]
\end{eqnarray}
The kinetic and exchange energy, $E_{t\,x}(\zeta=0)$,
of the unpolarized system,
shown in Fig.~\ref{figcones}(a), is obtained by
setting $k_F=0$ in Eq.~\ref{exeq}. The exchange energy
for the  case involving
{\it both} electron and hole carriers, Fig.~\ref{figcones}(c)
is given by a small change in Eq.~\ref{exeq}, where the
last term changes sign and becomes $-k_d^2\mathcal H_{14}(r)$.
If this is simplified and written using elliptic integrals,
as in Ref.~\onlinecite{pgcn},
the {\it change} in the exchange energy for electron and hole carriers,
compared to the unpolarized case becomes 
\begin{eqnarray}
\label{holexeq}
E_x(\zeta)&=&E_x(n_\delta,\zeta=0)+\Delta E_x(\zeta)\\
\Delta E_x(\zeta)&=&\frac{A_0}{(2\pi)^2}(g^0/4)(E_u/k_c)
\nonumber
[(k_u^3+k_d^3-2k_F^3)R_1(1)+\\
\nonumber
&&2k_ck_u^2R_2(k_u/k_c)-2k_ck_d^2R_2(k_d/k_c)\\
&&-4k_ck_F^2R_2(k_F/kc)]
\end{eqnarray}
As before, $E_u=\hbar V_Fk_c$ is the unit of energy.
For completeness, we note that
\begin{eqnarray}
E_x(n_\delta=0,\zeta=0)&=&-\frac{A_0}{(2\pi)^2}(g^0/2)E_uk_c^2R_1(1) \\
R_1(1)&=&3.776
\end{eqnarray}
Here the energy is per carbon atom
and we have used the notation $R_1(x), R_2(x)$ for the elliptic integrals
as given in Ref.~\onlinecite{pgcn}, and taken the opportunity to correct
a few typograpical errors in their Eq.~(15).  
 The energy difference which determines the competition between
 the unpolarized and polarized phases, i.e.,
$\Delta E_{k\,x}(\zeta)$, includes the kinetic-energy corrections
as well as the change in the exchange energy.
It is plotted in Fig.~\ref{figex}.
 We have done the calculations
in $r$-space using the PDFS, and in $k$-space via
elliptic integrals, to provide independent numerical checks. 
Fig.~\ref{figex} shows that stable spin-polarized phases appear in 
electron-carrier systems, if kinetic and exchange energies are used
in the total energy. A noteworthy feature of Fig.~\ref{figex} is the
strong cancellation of the kinetic energy by the exchange energy, leading to
net energies which are  about 1\% of the energy scale $E_u=V_Fk_c$.
Thus the stage is set for the phase stabilities to be determined by
the correlation energies which are left out in Fig.~\ref{figex}.
\begin {figure}
\includegraphics*[width=7.0cm, height=8.0cm]{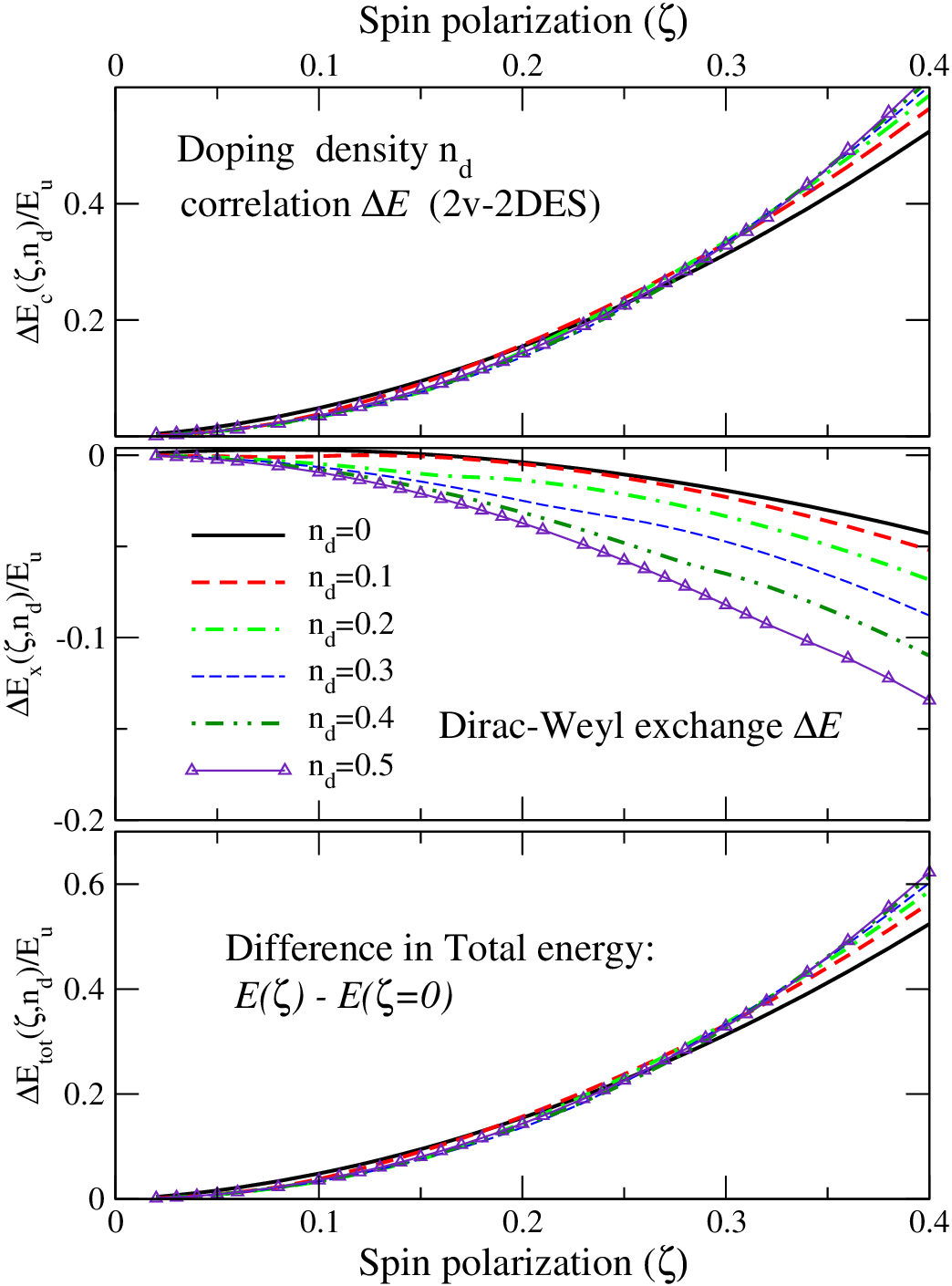}
\caption
{(Color online)Upper panel-
The correlation-energy difference $\Delta E_{c}(n_d,\zeta)$, i.e., 
between the polarized and
unpolarized phases, in units of $E_u=V_Fk_c$, as a function
of the spin polarization $\zeta$ and the dopent density $n_d$.
The lower panel shows the total energy difference between the
polarized and unpolarized phases.
The unpolarized state is the most stable phase.
}
\label{jelcorr}
\end{figure}
\section{Calculations including the correlation energy.} 
A rigorous calculation of the correlation energy requires the self-consistent
evaluation of $8\times8$ matrix of pair-distribution functions for many values 
of the coupling constant $0\le\lambda\le1$ in the interaction $\lambda/r$,
and an integration over the coupling constant $\lambda$.

 Although this can be envisaged
 within the CHNC approach, it still remains a very arduous task. Even when all the
 inherent symmetries in the problem are taken into account, some two-dozen PDFs
 need to be evaluated. Instead we outline a simplified scheme where we use the 
 results for the correlation energy of the two-valley, two-spin (i.e., 4
 component) jellium 2DES to reconstruct the correlation energy of the DW-2DES for
 equivalent values of the ratio of the Coulomb interaction and the kinetic
 energy, as discussed below. The method we use here provides a different, possibly
more transparent approach than that given in a previous
discussion\cite{sscdw}. The conclusions from the two methods are in agreement.

When the electron density of the jellium 2v-2DES is changed, the $r_s$ value changes.
In contrast, when the electron density (or, equivalently, the number of 
electrons per carbon) in the DW-2DES of perfect graphene is changed, the 
coupling constant $g^0$ remains unchanged. Hence the correlation energy
 per carbon for
a doping situation involving a total of $n_T$ electrons per carbon, at a
spin-polaraization $\zeta$ might be approximated by
 $n_T\epsilon^{2v}_c(r_s,\zeta)$, 
with $r_s=g^0$, where $\epsilon^{2v}_c(r_s,\zeta)$  is the corresponding
 2v-2DES correlation energy. 
However, this result is in terms of an
effective atomic unit $E_{au}$, intrinsic to the jellium-2DES, such that
 the Fermi energy $E_F=E_{eu}/r_s^2$. Since we identify the coupling constant
$g^0$ to be $r_s$, the effective atomic unit $E_{au}=E_Fr_s^2$.
A full evaluation of the 4-component jellium 2DES from the interacting
PDFs has been carried out and the correlation energy is given
 in parametrized form in Eq.~5 of
 Ref.~\onlinecite{2valley}.   
In transferring from the 2v-2DES to the DW-2DES we note that
$e^2/\epsilon_0$ which is unity in 2v-2DES becomes $g^0V_F$ in
the DW-2DES. The correlation energy enhancement is:
\begin{equation}
 \Delta E_c(n_d, \zeta)=E_c(n_d,\zeta)-E_c(n_d,0)
\end{equation}
The correlation energies in jellium-2DES are evaluated from PDFs whose
non-interacting forms are Bessel-type
PDFs, where as the DW-2DES contain only about 1/2 the number of Bessel-type PDFS,
 while the other are Struve-type PDFs which bringing in a minor contribution.
 Thus an upper bound would be to use the estimate of $\Delta E_c(n_d, \zeta)$
 given above, while a lower-bound would be about 1/2 the above estimate.
The $\Delta E_c$ calculated from the jellium 4-component system using
 the above scheme is
shown in Fig.~\ref{jelcorr}.  
The total energy difference inclusive of the kinetic energy, 
$E_x$, and the estimated $E_c$, 
between the polarized and
unpolarized phases is shown in in the lower panel of 
Fig.~\ref {jelcorr}. Sice the kinetic-energy enhancement of the
polarized phase is compensated by the exchange enhancement, the
total energy enhancement is determined by the correlation
effects. 
Thus we see that the inclusion of
the correlation energy suppresses the spin-polarized phase
found in the exchange-only calculation.

In this work we have kept the Coulomb
coupling fixed at $g^0=2.7$ typical of graphene,
unlike in other studies\cite{tosatti,pgcn} where
the coupling strength $g$ is taken as a tunable parameter,
in the spirit of Hubbard-model studies. If dielectric screening
is taken into account\cite{bedell}, the coupling is reduced
and many-body effects become smaller. We do not see a
practical experimental scheme for increasing the value of the
Coulomb coupling strength $g^0$ beyond 2.7 in the graphene system.

Even in the one-valley 2DES, the SPP of low-order theories is
pushed to $g\sim 26-27$. In the 2v-2DES,
direct terms predominate over
exchange interactions, and the SPP is not found in
CHNC\cite{2valley} or QMC\cite{conti} calculations.
We see that the inclusion of correlations
within a reasonable scheme suppresses the exchange-driven SPP
in the graphene-2DES as well. This result is in agreement with the
conclusions of Ref.~\onlinecite{sscdw}.

\end{document}